# About the influence of the temperature and pressure on the fractal dimension of defects in crystal structures

© **Ikhtier Holmamatovich Umirzakov**
*The laboratory of modeling. Kutateladze Institute of Thermophysics of SB of RAS.*
*Prospect Lavrenteva, 1. Novosibirsk 630090 Russia.*
*Tel.: +7 (383) 354-20-17. E-mail: tepliza@academ.org.*

***Keywords***: *defects in crystal, dislocation loop, distribution of defects.*

**Abstract**
It is shown that the results presented in the article "The influence of the temperature and pressure on a fractal dimension of defects in crystal structures" by V.B. Fedoseev (Butlerov communications. 2010. V23. №14. P.36-42) are not correct, because the results are based on impracticable conditions, conflicting assumptions and incorrect distributions.

# К вопросу о влиянии температуры и давления на фрактальную размерность дефектов кристаллической структуры

© Умирзаков Ихтиёр Холмаматович

*Лаборатория моделирования, Институт теплофизики СО РАН,
проспект Лаврентьева, 1, 630090, г. Новосибирск, Россия
Тел.: (383) 354-20-17, E-mail: tepliza@academ.org*

## Аннотация

Показано, что результаты, полученные в статье Федосеева В.Б. «Влияние температуры и давления на фрактальную размерность дефектов кристаллической структуры» (*Бутлеровские сообщения.* **2010**. Т.23. №14. С.36-42.) недостоверны, так как основаны на невыполнимых условиях, противоречивых предположениях и неверных распределениях.

## Введение

В настоящей работе показано, что исходные предположения автора [1] противоречивы и неверны, используемые им распределения неправильны, поэтому все результаты работы недостоверны.

## Основная часть

1. Согласно формуле (2) из [1]

$$N_v = v \cdot n_v, \tag{a}$$

где $v$ - стехиометрическое число дефекта, равное минимальному количеству атомов, перемещением которых в идеальной кристаллической решетке можно получить соответствующий дефект, $n_v$ - число дефектов со стехиометрическим числом $v$, $N_v$ - число атомов, перемещением которых в идеальной кристаллической решетке можно получить $n_v$ дефектов.

Из (3) в [1] с использованием указанного выше соотношения (a) имеем

$$N_v = \frac{N}{N-v} \cdot \exp(-u_v/RT) \cdot \exp[\pi \cdot (\sqrt{2(N-v)/3} - \sqrt{2N/3}]/A, \tag{b}$$

где $N$ - число атомов в рассматриваемом монокристаллическом блоке, $T$ – температура, $R$ – универсальная газовая постоянная, $A$ – нормирующий множитель, $u_v$ - средняя энергия, приходящаяся на один атом, участвующий в образовании дефекта со стехиометрическим числом $v$.

Очевидно, что нормирующий множитель $A$ не должен зависеть от $v$.

Из формулы без номера, находящейся перед формулой (3) в [1] имеем

$$N_v = N \cdot \exp(-u_v/RT) / \sum_v \exp(-u_v/RT). \qquad (c)$$

Из формул (b) и (c) имеем

$$A = \frac{1}{N-v} \cdot \exp[\pi \cdot (\sqrt{2(N-v)/3} - \sqrt{2N/3}] \cdot \sum_v \exp[-u_v/kT]. \qquad (d)$$

Из этой формулы следует, что нормирующий множитель $A$ зависит от $v$. Указанное противоречие говорит о неверности формулы (3) в [1] (формула (b)) и/или формулы, находящейся до формулы (3) в [1] – (формула (c)). Согласно [1] формула (c) представляет собой каноническое распределение Гиббса, а формула (3) получена минимизацией функции Гиббса. В работе [1] процедура минимизации не указана в явном виде, поэтому только можно предположить, что формула (b) была получена с использованием формулы (c). В этом случае, если минимизация проведена правильно, то неверны обе формулы (b) и (c).

Работы [1,3-6] автора [1] не содержат вывода формулы (c). Т.е. автор [1,3-6] необоснованно использует формулу (c). Работы [1,3-5] не содержат также вывода формулы (b). «Вывод» формулы (b) в [6] неверен, так как в формуле (5) в [6] не учтена возможность наличия двух и более полимеров одного и того же размера, а также полимеров других размеров. Формула (5) в [6] неверно отождествлена с вероятностью присутствия полимера p в системе. Это следует из того, что вероятность образования полимера любого размера, равная сумме этих вероятностей, оказывается числом больше чем 1, что абсурдно. Причина этого заключается в том, что автор [6] не учел возможность наличия в системе множества полимеров с разными размерами. Таким образом, формула (c) необоснованна и неверна, и используется в работах [1,5,6].

2. Согласно (1) из [1]

$$N = \sum_v v \cdot n_v. \qquad (e)$$

где N – число атомов в кристалле. Следовательно, нет атомов, оставшихся неподвижными в идеальной кристаллической решетке, нет атомов, находящихся между дефектами, и в образовании дефектов участвуют все атомы кристалла. Поэтому все дефекты взаимодействуют между собой прямо или косвенно. В таком случае не может идти речь об отдельных дефектах. То есть весь кристалл представляет собой один макроскопический дефект. Следовательно, условие сохранения энергии (2) в [1] в приближении идеального раствора, не учитывающего энергию взаимодействия дефектов (компонентов) между собой, неверно. Это означает, результаты [1], полученные с использованием формул (1) и (2) из [1] неверны. Равновесное распределение дефектов (формула (3) в [1]) получено на основе формул (1) и (2) из [1]. Следовательно, все графики и выводы в [1] полученные на основе формулы (3) и формулы (8), полученной на основе (3), недостоверны.

3. Формула (1) из [1] неверна в общем случае, так как в кристалле не всегда

существует много дефектов, которые взаимодействуют между собой прямо и косвенно. При низких температурах в кристалле мало дефектов и они не взаимодействуют между собой, и не все атомы кристалла участвуют в образовании дефектов.

Кроме того, в формуле (1) никак не учтено, что один и тот же атом может участвовать при образовании разных дефектов.

Эта же ошибочная формула (1) используется также в работах [2-6] автора [1].

4. Условие сохранения суммарной энергии дефектов неверно с физической точки зрения, так как нет никаких физических или иных способов для поддержания постоянной этой энергии. Только энергию кристалла можно поддерживать постоянной.

Это же ошибочное условие используется также в работах [2-6] автора [1].

5. Если даже предположить, что условие сохранения суммарной энергии дефектов выполняется, то условиям сохранения (1) и (2) отвечает микроканоническое распределение [2], а не каноническое распределение Гиббса, как утверждается в [1].

6. В формуле для канонического распределения, находящейся между формулами (2) и (3) в [1] (формула (c)) никак не учтено условие сохранения энергии (2), поскольку она никак не зависит от суммарной энергии дефектов $U$.

7. Формула для канонического распределения, находящаяся между формулами (2) и (3) в [1] (формула (c)) неверна, так как каноническое распределение зависит от свободной энергии дефектов, а не от энергии $u_v$.

**Выводы**

1. Предположение работы [1] об участии в образовании дефектов всех атомов кристалла неверно в общем случае.
2. Предположение работы [1] о сохранении суммарной энергии дефектов в кристалле физически невыполнимо и поэтому неверно.
3. В работе [1] для описания ансамбля дефектов с условиями сохранения полного стехиометрического числа (формула (1) в [1]) и сохранения энергии дефектов (формула (2) в [1]) неправильно используется каноническое распределение Гиббса вместо микроканонического распределения и неправильно применена минимизация функции Гиббса вместо условия максимума энтропии.
4. Полученные в [1] распределения противоречат друг другу, что свидетельствуют об ошибках, сделанных автором [1] при выводе этих распределений.
5. Все выводы работы [1], основанные на этих распределениях, недостоверны.
6. Статья представляет собой на первый взгляд работу, основанных на разумных предположениях, а реально при детальном анализе оказывается, что она представляет собой в основном набор логически несвязанных между собой различных предложений и наборов предложений.
7. Неверны результаты работ [2-6], полученные с использованием условий (1), (2) из [1] и формул (b) и (c).

**Литература**